# Warrant Theory

**Khashayar Irani**
**Birkbeck College, University of London**
**k.irani@mathematicallogic.com**
**September 2026**

**Abstract**

In this paper, we develop warrant theory as a philosophical discipline concerned with the inferential legitimacy of propositions within logical analysis.[1] Warrant theory reconceptualises logic as a normative framework governing the conditions under which propositions may be introduced, accepted, rejected, and inferentially employed.[2] Warrant is understood as inferential entitlement and is distinguished from truth, belief, and other psychological attitudes, while its relation to inferential use and meaning is examined.[3] Warrant-theoretic analysis is then developed as a systematic method for investigating how propositions acquire inferential standing, how that standing develops, and how inferential positions interact through relations of dependence, compatibility, incompatibility, and exclusion. Acceptance and rejection provide the bilateral vocabulary for representing positive and negative inferential positions and the consequences and commitments associated with them. Finally, these elements are brought together in a warrant-theoretic definition of logic as the formal and normative study of the conditions under which propositions may be legitimately accepted or rejected and of the inferential transitions that such legitimacy warrants. On this account, logical consequence and logical failure are understood through the presence, preservation, or absence of inferential entitlement, thus locating the philosophical subject matter of logic in the systematic governance of inferential legitimacy.

[1] In our view, warrant theory can be a very rich philosophical discipline that may be employed to address many philosophical questions in the philosophy of logic, philosophy of language, epistemology, ethics, and beyond. Although we cannot cover all areas of warrant theory in this manuscript, we nevertheless do our best to lay down the foundations for subsequent warrant-theoretic studies. Furthermore, in our formulation of warrant theory, in order to fully capture the aims we pursue, we present our theory within a bilateral framework rather than a purely unilateral one. As a result, we analyse logical propositions as possessing positive and negative force. Moreover, at the beginning of this work, we should emphasise that by the term propositions, we sometimes mean well-formed formulae in a formal language or, on other occasions, declarative statements in natural language. But we do not attach any metaphysical attribute whatsoever to the concept of proposition, and our use of this notion is purely linguistic. Finally, throughout this text, by logical analysis we mean the systematic analysis of reasoning and language through the formal techniques of logic, including the examination of propositions, their inferential relations, and the rules governing transitions between them. The term is therefore used broadly to encompass both formal reasoning and the logical analysis of linguistic expressions, without implying that logical analysis is reducible to either syntactic manipulation or semantic evaluation alone.

[2] For a general, but detailed discussion concerning the notion of the normative status of logic, see Florian Steinberger's excellent (2022) article and the references cited therein.

[3] As we shall see, the notion of inferential entitlement that we discuss here is similar to, but not identical with, the concept of inferentialism found in logical inferentialism. For an introduction to the theory of logical inferentialism and the challenges it poses to classical logic, see Khashayar Irani (2025). However, for a detailed examination and formulation of logical inferentialism, see Steinberger (2009).

## 1. What is Warrant Theory?

Warrant theory begins from a fundamental reconsideration of what logical analysis is meant to explain. Rather than restricting itself to questions of truth, falsity, or formal validity, it asks a prior and more foundational question that is to say, what entitles a proposition to function within logical analysis at all? Before a statement can serve as a premise, a conclusion, or a step within a proof, it must possess a form of inferential legitimacy. Warrant theory is concerned with identifying and analysing this legitimacy. It therefore shifts the focus of logic from mere evaluation to admissibility, from truth alone to inferential entitlement, and from formal relations to normative structure. In doing so, it presents logic not simply as a system of truth-preserving relations, but as a structured space in which propositions are warranted, positioned, and governed. This shift is philosophically significant because traditional approaches, especially those grounded in truth-conditional semantics, leave an important dimension of logic unexplained. While they can determine whether an argument is valid or whether a proposition is true under an interpretation, they do not fully account for what makes a proposition properly available for inferential use. A formula may be syntactically well-formed and semantically interpretable, yet this alone does not explain why it may legitimately enter into the space of inferential analysis. Warrant theory addresses this gap by introducing the notion of warrant as a condition of inferential admissibility. It asks not only whether propositions can be evaluated, but under what conditions they may be accepted or rejected within an inferential framework. A prominent advocate of the notion of warrant is Michael Dummett (1973), who built his theory of meaning around this view. He states:

> Crudely expressed, there are always two aspects of the use of a given form of sentence: the conditions under which an utterance of that sentence is appropriate, which include, in the case of an assertoric sentence, what counts as an acceptable ground for asserting it; and the consequences of an utterance of it, which comprise both what the speaker commits himself to by the utterance and the appropriate response on the part of the hearer, including, in the case of assertion, what he is entitled to infer from it if he accepts it.[4]

In this passage, Dummett's notion of the "two aspects of use" of a sentence refers to the concept of warrant. It indicates, first, the conditions under which one is entitled to accept a statement, and second, the inferential consequences that such acceptance commits one to within the structure of logical analysis.[5] Another renowned philosopher who writes about the notion of warrant is David Ripley in his (2017) paper *Bilateralism, Coherence, Warrant*.

---

[4] Dummett (1973) Page: 396

[5] Of course, in the given quote, Dummett (1973) specifically speaks in terms of "assertion", but we reinterpret this as "acceptance" in order to maintain consistency with our use of that concept throughout this work.

Ripley presents the concept of warrant as the orthodox epistemic basis on which many theories of meaning have traditionally relied. He states:

> The orthodox view is that the conditions invoked by both unilateralists and bilateralists are conditions under which an assertion or denial is warranted. On this warrant-based account, a sentence's assertion conditions are the conditions under which it may be warrantedly asserted; for bilateralists, its denial conditions are in addition the conditions under which it may be warrantedly denied. Semantics, on this view, is at root epistemological; it is a matter of justification, whatever justification in the end itself amounts to. This warrant-based conception is common to many unilateralists and bilateralists alike (witness, for example, Dummett 1991, Price 1983, Rumfitt 2000, and Tennant 1987).[6]

However, in addition to Dummett (1973 & 1991), Huw Price (1983), Neil Tennant (1987), Ian Rumfitt (2000), Greg Restall (2004), Ripley (2017), and other authors who have written extensively on the notion of warrant, one figure who must not be forgotten as a systematic initiator of the concept is Gerhard Gentzen. In his celebrated (1935) proof-theoretic work, Gentzen writes:

> The introductions represent, as it were, the 'definitions' of the symbols concerned, and the eliminations are no more, in the final analysis, than the consequences of these definitions. This fact may be expressed as follows: In eliminating a symbol, we may use the formula with whose terminal symbol we are dealing only 'in the sense afforded it by the introduction of that symbol.[7]

The above passage may be understood, within the present framework, as an early and precise articulation of the notion of warrant in proof-theoretic terms. Gentzen's claim that introduction rules function as the "definitions" of logical symbols namely, logical constants, can be interpreted as identifying the conditions under which a proposition is warranted for acceptance within an inferential analysis. To introduce a constant is not simply to form a new expression, but to establish the inferential conditions that warrant its legitimate use. In this sense, the introduction rules determine the grounds upon which a proposition involving a given constant may be accepted, and thus they fix its inferential standing. The corresponding elimination rules, then, do not introduce new content independently, but articulate what one is entitled to infer from a proposition once it has been warranted through its introduction. This reflects precisely the two aspects of warrant already discussed by authors like Dummett (1973), namely, the conditions of acceptance and the consequences that follow from such acceptance.

Gentzen's further claim that a formula may be used in elimination only "in the sense afforded" by its introduction reinforces this point. It indicates that the inferential use of a proposition must remain governed by the very conditions that warranted it in the first place. In warrant-theoretic terms, this means that inferential transitions are legitimate only insofar as they preserve the entitlement established by the introduction rules. Accordingly, Gentzen's framework can be read as providing a systematic account of how propositions acquire and transmit warrant within logical analysis, where the meaning and use of logical constants are inseparable from the inferential conditions that justify their acceptance and regulate their employment. To clarify Gentzen's argument concerning the introduction and elimination rules of logical constants and their ability to generate warrant, consider the following rules ∧I and ∧E:[8]

---

[6] Ripley (2017) cited in Friederike Moltmann & Mark Textor (2017) Page: 309. Of course one might argue that Ripley's central aim is to show that warrant is ultimately not the right foundation for semantics. Nonetheless, his contribution to this subject cannot go unnoticed.

[7] Gentzen (1935) cited in M. E. Szabo (1969) Page: 80.

$$\frac{\mathbf{A} \qquad \mathbf{B}}{\mathbf{A} \wedge \mathbf{B}}\wedge\mathbf{I} \qquad\qquad \frac{\mathbf{A} \wedge \mathbf{B}}{\mathbf{A}}\wedge\mathbf{E} \qquad\qquad \frac{\mathbf{A} \wedge \mathbf{B}}{\mathbf{B}}\wedge\mathbf{E}$$

Within a warrant-theoretic framework, these rules may be understood as providing a clear illustration of how warrant is both generated and transmitted through logical constants. The rule ∧I specifies the precise inferential conditions under which a conjunctive proposition may be accepted. If proposition A is already warranted ($\Gamma \Rightarrow A$) for acceptance, and proposition B is likewise warranted ($\Gamma \Rightarrow B$) for acceptance, then one becomes inferentially entitled to accept the compound proposition $A \wedge B$. Therefore, the rule does not simply permit the syntactic formation of a conjunction; rather, it establishes the grounds upon which the conjunction acquires warrant. The acceptance of $A \wedge B$ is thus justified entirely by the prior warrant attached to its constituent propositions. In Gentzen's terminology, this introduction rule effectively defines the meaning of conjunction, because it determines the conditions under which a conjunctive proposition may legitimately enter logical analysis. From the standpoint of warrant theory, this means that conjunction receives its positive inferential standing through the warrant transmitted from both conjuncts. Conversely, the rule ∧E governs what one is entitled to infer once the conjunction has already been warranted. If $A \wedge B$ has been legitimately accepted, then one is entitled to recover A, and likewise entitled to recover B. These elimination rules do not generate new warrant independently; rather, they unpack or transmit the warrant already contained within the accepted conjunction. This is precisely what Gentzen means when he states that elimination must use a formula only in the sense afforded by its introduction. Since the conjunction was introduced through the joint warrant of A and B, its legitimate inferential use must preserve that structure. Accordingly, in warrant theory, the harmony between ∧I and ∧E:

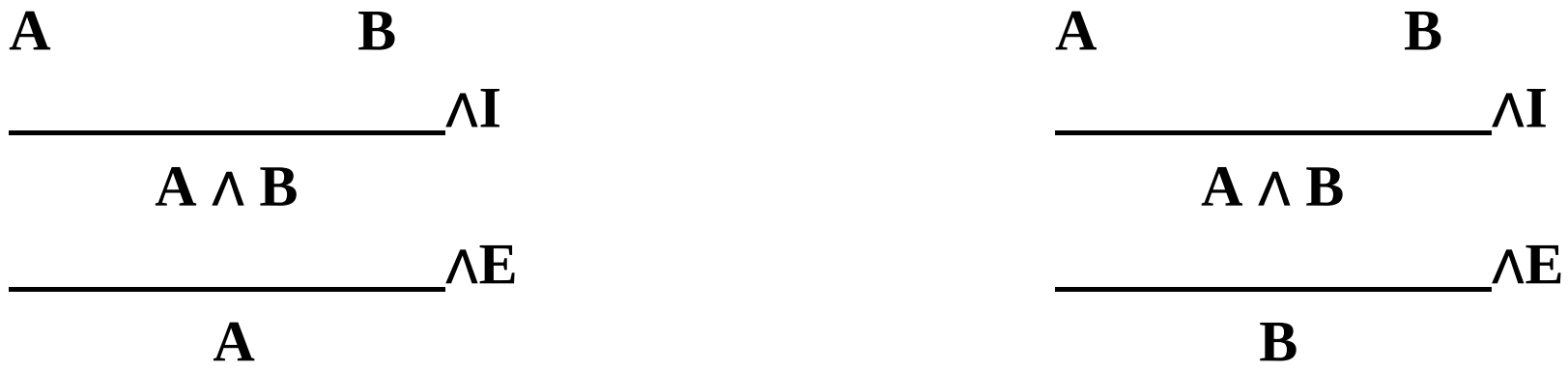

[8] For a comprehensive study concerning Gentzen's system of natural deduction, see Dag Prawitz (1965 & 1971).

demonstrates that logical constants regulate not only the formation of propositions but also the disciplined movement of warrant through logical analysis, thus making inferential legitimacy both structured and intelligible.[9]

Furthermore, at its core, warrant may be defined as the inferential entitlement that licenses a proposition to function within logical analysis. It concerns the legitimacy of a proposition's place within analysis. To declare that a proposition is warranted is not simply to say that it is meaningful, nor even simply that it is true, but that it is authorised for inferential use. Consequently, warrant belongs to the normative dimension of logic since it governs what one is entitled to maintain, what one may infer, and what must be excluded. This can be illustrated through familiar forms of disciplined analysis. In a court of law, for instance, a juror may strongly believe that a defendant is guilty based on personal impressions or prejudice; however, such belief does not warrant the conclusion unless it is supported by admissible evidence and properly structured inference. Conversely, a verdict may be warranted on the basis of conclusive forensic evidence and testimony, even if an individual juror feels uncertain or emotionally unconvinced. A similar pattern appears in scientific practice. A researcher may believe that a hypothesis is correct because it appears intuitively plausible, yet without experimental confirmation and reproducible results, the hypothesis is not warranted. On the other hand, a scientific claim may be fully warranted through rigorous data and established methodology, even if the researcher personally doubts or hesitates to accept it. These examples demonstrate the central claim that a proposition may be believed without being warranted, and it may be warranted without depending upon any particular mental state. In this sense, warrant theory is not simply an extension of semantics or proof theory, but a framework that explains how propositions acquire their inferential standing in the first place. As a result, a crucial feature of this account is the sharp distinction between warrant and psychological or pragmatic notions such as belief, assertion, or conviction. Warrant is thus objective in the sense that it belongs to the formal and public structure of logical analysis rather than to subjective attitudes, and this allows warrant theory to avoid psychologism while preserving a rigorous account of logical legitimacy.[10]

---

[9] Note that in warrant-theoretic terms, the harmony between $\wedge$I and $\wedge$E concerns the preservation of inferential entitlement across the introduction and subsequent use of a conjunction. The rule $\wedge$I establishes the conditions under which $A \wedge B$ acquires warrant because if A and B are independently warranted for acceptance, their joint inferential standing warrants the acceptance of $A \wedge B$. The conjunction therefore possesses no inferential entitlement beyond that grounded in the warrant of its constituent propositions. The elimination rule $\wedge$E operates in the opposite direction by determining what one is entitled to recover from the warranted conjunction. Since $A \wedge B$ was introduced on the basis of warrants for both A and B, its subsequent use legitimately permits the recovery of either conjunct. Harmony consists precisely in this correspondence between the conditions governing introduction and the consequences licensed by elimination. $\wedge$E neither extracts information unsupported by $\wedge$I nor introduces an independent source of warrant; rather, it preserves and transmits the entitlement already incorporated into the conjunction. Consequently, the introduction and elimination rules jointly regulate the inferential life of $A \wedge B$ since $\wedge$I determines how warrant enters the conjunctive position, while $\wedge$E determines how that warrant may legitimately flow from it. Thus, harmony expresses a disciplined preservation of inferential entitlement.

[10] With respect to the notion of psychologism, Martin Kusch (2024) writes: *'Psychologism' entered the English language as a translation of the German word 'Psychologismus', a term coined by the Hegelian Johann Eduard*

Equally important is the distinction between warrant and truth. Although related, they are not identical. Truth concerns what is the case, whereas warrant concerns what is inferentially admissible within a disciplined framework of analysis. A proposition may therefore be true without being warranted in a given inferential context, and, conversely, a proposition may be warranted as part of a properly governed systematic process without that warrant being reducible to a direct correspondence with reality. This can be illustrated through both ordinary and formal examples. Suppose, for instance, that a patient does in fact have an underlying illness at an early stage, yet the doctor possesses no test results, no reliable symptoms, and no medically sufficient grounds for diagnosis. In such a case, the statement 'The patient has the illness' may be true, but it lacks warrant and therefore cannot function meaningfully within medical examination, since it does not warrant further justified inference. Similarly, a witness may truthfully state that an event occurred, yet without corroboration or evidential grounding, that statement cannot be used inferentially within a legal argument, and thus fails to acquire full methodical significance. By contrast, within formal analysis, inferentially warranted propositions acquire meaning precisely through their role in warranted transitions. For example, as was previously stated, in natural deduction, from A and B one may infer $A \wedge B$ by $\wedge$I, and from $A \wedge B$ one may infer A by $\wedge$E. These transitions do not just preserve truth; they determine how the propositions are to be understood within logical analysis. A proposition such as $A \wedge B$ is meaningful not simply because it may be true under some interpretation, but because it stands within a network of warranted inferential uses namely, that it may be introduced from A and B and may warrant the recovery of each conjunct. Likewise, a conditional similar to $A \rightarrow B$ is not understood purely as a truth-functional relation, but through its warranted use; that is to say, from the already derived B ($\Gamma \Rightarrow B$) and the presence of the assumption [A], one may infer the conditional $A \rightarrow B$.[11] These examples show that meaning is inseparable from warranted use. A purely true proposition, if inferentially inert or unwarranted, fails to participate in the structure of logical analysis and thus remains methodically incomplete. It is only when a proposition is inferentially warranted, when it can be introduced, sustained, and employed within a system of warranted transitions that it acquires full logical significance. This distinction enables warrant theory to supplement truth-based approaches by revealing a dimension of logic that concerns inferential authorisation rather than semantic valuation. As a result, warrant theory provides a deeper account of logical structure by explaining how propositions become analytically usable within an inferential system. This reinterpretation also has important consequences for the theory of meaning. To understand a proposition is not purely to know the conditions under which it is true, but to grasp the conditions under which it may be accepted, the conditions under which it must be rejected, and the inferential consequences that follow from each. Hence, meaning is inseparable from warranted use. A proposition acquires significance not only by what it represents, but by how it functions

*Erdmann in 1870 to critically characterize the philosophical position of Eduard Beneke (Erdmann 1870). Although the term continues to be used today, criticisms and defenses of psychologism have mostly been absorbed into wider debates over the pros and cons of philosophical naturalism.* For a more detailed investigation, see Kusch's (2024) article on this matter.

[11] As always, formulae shown inside square brackets [ ] are assumed formulae, or assumptions.

within logical analysis. Therefore, and speaking systematically, warrant theory proposes that meaning is the inferential consequence of a proposition, determined by the conditions under which it may be legitimately accepted or rejected, and by the network of warranted inferences and exclusions in which it is entitled to participate within logical analysis.[12] In this mode, warrant theory offers a unified perspective on logic, meaning, and rationality, revealing that logical analysis is fundamentally concerned with inferential legitimacy, where propositions are admitted, positioned, and governed within a structured normative space, and where logic itself concerns not simply what follows from what, but what is entitled to follow and why.

## 2. What is Warrant-Theoretic Analysis?

If the preceding discussion has established warrant as the inferential entitlement by virtue of which a proposition may function within logical analysis, warrant-theoretic analysis concerns the systematic procedure by which the operation of that entitlement is examined. Its task is therefore methodological rather than definitional. It does not ask again what warrant is, nor does it repeat the distinction between warrant, truth, belief, or psychological conviction. Instead, it begins from the availability of warrant as an analytical category and investigates how inferential standing is acquired, maintained, transferred, restricted, or lost within a structured course of analysis. The central object of warrant-theoretic analysis is consequently not the proposition considered in isolation, but the proposition as situated within an inferential environment. Such analysis asks what grounds permit a proposition to enter that environment, what role it is entitled to occupy once introduced, and what consequences arise from its continued employment. This gives warrant-theoretic analysis a distinctive direction since it proceeds from conditions of entry to conditions of use, and from conditions of use to the consequences and restrictions generated by inferential participation. A proposition may therefore be examined at successive stages of its analytical career without requiring a return to the more general philosophical question of warrant itself. The method is concerned with the organisation of inferential standing across an analysis and with the relations through which that standing becomes operative. In this respect, warrant-theoretic analysis converts the general notion of entitlement developed in the preceding section into a precise instrument of investigation. It asks, in each case, not simply whether a proposition appears within an argument or derivation, but whether its presence has an identifiable inferential basis, whether its subsequent employment remains continuous with that basis, and whether the positions generated from it remain admissible within the inferential structure presently under examination.

The first principal task of warrant-theoretic analysis is to determine the conditions under which a proposition acquires a determinate inferential position and becomes available for systematic use. Introduction into an analysis cannot be identified purely with inscription, formulation, or occurrence. A proposition may occur without thus possessing the standing required for inferential employment. Warrant-theoretic analysis therefore distinguishes the presence of a proposition from its admissible presence. The relevant question is what grounds its occupation of a particular position within the analysis and what that position permits. Consider, for example, an historical investigation in which a discovered letter states that a treaty was negotiated secretly before its publicly recorded date. The sentence may be intelligible and relevant to the inquiry, yet its mere appearance in the archive does not determine the inferential role it may occupy. Its provenance, date, authorship, relation to other records, and evidential dependencies must first be established before conclusions may properly depend upon it. Warrant-theoretic analysis abstracts from the subject matter of this example and concentrates upon the structure thus exhibited for the reason that a proposition enters analysis under determinate conditions, depends upon specified grounds, and acquires a position whose subsequent use is constrained by those grounds. The same attention to provenance applies within formal analysis, where a formula may occur as a premise, temporary assumption, intermediate result, or derived conclusion. These modes of occurrence are not inferentially interchangeable. The method therefore records where a proposition enters an analysis, upon which prior positions it depends, what restrictions accompany its introduction, and whether those restrictions continue to govern its later employment. Two occurrences of the same formula may consequently possess different analytical statuses when their inferential histories differ. By reconstructing those histories, warrant-theoretic analysis provides a disciplined account of inferential availability and prevents mere occurrence from being mistaken for warranted participation.

[12] For future research reference, we shall label the meaning theory that arises from the discipline of warrant theory as warrant-theoretic semantics.

The second principal task of warrant-theoretic analysis is to examine what happens to inferential standing as logical analysis develops. Once a proposition has acquired an admissible position, that position may support further transitions, interact with other established positions, or become subject to restrictions introduced at later stages. The method therefore tracks inferential dependency across an analysis and asks whether each subsequent employment remains continuous with the grounds upon which the relevant proposition became available. A formal illustration is disjunction elimination:

$$\dfrac{\mathbf{A \lor B} \qquad \begin{matrix}\mathbf{[A]} \\ \mathbf{\Pi} \\ \mathbf{C}\end{matrix} \qquad \begin{matrix}\mathbf{[B]} \\ \mathbf{\Pi} \\ \mathbf{C}\end{matrix}}{\mathbf{C}} \lor\mathbf{E}$$

Suppose A ∨ B has been established, C is derived from the temporary assumption [A], and C is also derived from the temporary assumption [B]. The eventual derivation of C depends upon a structured coordination of these separate inferential routes. Warrant-theoretic analysis does not purely register the final formula C or note that ∨E has been applied. It reconstructs the dependencies through which C becomes available, distinguishes the temporary assumptions from propositions available outside their subderivations, and examines whether the discharge of those assumptions leaves C with the required standing. What matters is continuity through inferential transformation. This illustrates the diagnostic character of the method. If one branch fails to establish C, if an assumption is employed beyond its permitted scope, or if the conclusion depends upon a position that has ceased to be available, the analysis identifies where the inferential chain becomes defective. Conversely, where the dependencies are properly maintained, the method displays how inferential standing survives changes in the structure of a derivation. Warrant-theoretic analysis therefore treats logical analysis as an organised sequence of dependent positions rather than as a succession of formulas connected only by rule labels. Its concern is to reconstruct the path of inferential authority, determine where that path remains intact, and identify precisely where an attempted transition ceases to possess the conditions required for admissible continuation.

A further task of warrant-theoretic analysis is to examine inferential positions relationally, since the standing of a proposition is revealed partly through the effects its presence has upon the surrounding analytical configuration. Once positioned, a proposition may strengthen one route of inquiry, render another unavailable, remain compatible with some propositions, or generate tension with alternatives. Consider an archaeological investigation in which carbon dating places an artefact within a century while an inscription initially appears to assign it to a later period. Warrant-theoretic analysis need not decide the archaeological question. Rather, it examines the inferential configuration produced by the two propositions namely, what each position depends upon, whether they can be jointly sustained, what auxiliary propositions might reconcile them, and what consequences follow if one position is withdrawn or restricted. The example illustrates that inferential standing cannot always be analysed atomistically. Its operation becomes visible through relations of dependence, compatibility, incompatibility, and exclusion. Accordingly, warrant-theoretic analysis maps not only individual inferential histories but also the field formed by their interaction. This relational dimension completes the methodological picture developed in this section. Analysis first identifies how propositions become available, then traces how their standing is preserved or transformed, and finally examines how established positions constrain the wider inferential field. At this point, however, the method encounters a distinction that cannot be expressed simply by saying that propositions possess inferential standing. An adequate analysis must also distinguish the direction in which a proposition is positioned that is, whether it is sustained within the inferential field or excluded from it. That distinction introduces no new account of warrant; rather, it supplies the vocabulary required to represent the opposed orientations disclosed by warrant-theoretic analysis itself. The next section therefore turns to acceptance and rejection as the formal indicators through which positive and negative inferential positions can be articulated.

## 3. The Role of Acceptance & Rejection in Warrant Theory

If warrant theory identifies the inferential entitlement by virtue of which propositions may function within logical analysis, and warrant-theoretic analysis examines the acquisition, development, and interaction of that entitlement, the notions of acceptance and rejection provide the bilateral vocabulary required to distinguish the direction of the inferential positions thus disclosed. As the preceding section indicated, determining that a proposition possesses inferential standing does not yet specify whether that standing is positive or negative, that is, whether the proposition is sustained within the inferential field or excluded from it. Acceptance and rejection articulate precisely this distinction. Within warrant theory, to accept a proposition is to assign it a positive inferential position, whereas to reject a proposition is to assign it a negative inferential position.[13] These notions therefore do not introduce a further kind of warrant in addition to the inferential entitlement already established, but specify the opposed orientations that warranted positioning may assume within a bilateral framework. Their significance lies in making the polarity of inferential structure explicit because warrant-theoretic analysis can thus represent not only that a proposition occupies a determinate position, but also the direction in which that position operates within logical analysis. Acceptance and rejection should consequently be understood as formal indicators of inferential orientation rather than as psychological attitudes or descriptions of linguistic behaviour. Although the vocabulary may resemble ordinary acts of accepting, asserting, rejecting, or denying statements, its function within warrant theory is specifically inferential, since it concerns the positive or negative position occupied by a proposition rather than the mental state or communicative performance of a speaker. In each case, acceptance and rejection function as indicators of inferential legitimacy rather than expressions of inner conviction or outward assertion. They are not part of a theory of speech acts, nor do they constitute a system of illocutionary logic, but serve instead to articulate the structured positioning of propositions within logical analysis.[14] However, prior to continuing with our discussion concerning the role of acceptance and rejection in warrant theory, it should be noted that a classical example of a system of illocutionary logic is John Rogers Searle's and Daniel Vanderveken's (1985) book *Foundations of Illocutionary Logic*, in which their main project is to study the logical relations among speech acts. They elucidate their primary aim as:

> A theory of illocutionary logic of the sort we are describing is essentially a theory of illocutionary commitment as determined by illocutionary force. The single most important question it must answer is this: Given that a speaker in a certain context of utterance performs a successful illocutionary act of a certain form, what other illocutions does the performance of that act commit him to?[15]

Nevertheless, this distinction must be maintained throughout the present account, since conflating inferential positions with illocutionary acts would obscure the specifically bilateral function of acceptance and rejection in warrant theory. It is this bilateral function, and the distinct inferential consequences associated with each orientation, that now requires further examination.

---

[13] If we wish to represent a proposition, for example A, within a formal environment through acceptance and rejection, then, bilaterally speaking, we assign a plus sign to A, represented as +A, to denote the acceptance of A, and a minus sign to A, represented as -A, to denote the rejection of A. It should be noted, however, that in a formal setting the signs + and - are not embedded within A itself; rather, they belong to the structural segment of our formal language and indicate the inferential position assigned to A. This distinguishes them from the negation operator ¬, which is embedded within the formula itself. Thus, +A and -A represent, respectively, the positive and negative inferential positions assigned to the same proposition A, whereas ¬A constitutes a distinct formula formed by applying the negation operator to A. Accordingly, so long as we have assigned neither + nor - to ¬A, and thus have obtained neither +(¬A) nor -(¬A), we have not yet specified our warrant-theoretic position towards ¬A. For a further examination of the bilateral role of + and - within an inferential environment, see Rumfitt (2000).

[14] It may be helpful to emphasise that, the subject matter of illocutionary logic includes issues such as descriptions, interviews, deliberations, consultations, regulations, evaluations, protestations, and eulogies.

[15] Searle & Vanderveken (1985) Page: 6

The importance of acceptance and rejection lies in the distinct inferential consequences associated with the positions they express. Once a proposition is accepted, its positive position may warrant further transitions, support conclusions, and constrain which other propositions remain available within the analysis. Rejection operates differently since assigning a proposition a negative position may prevent inferential routes that depend upon its acceptance, exclude certain possibilities, and alter which positions remain available. Consider, for example, an investigation in which the proposition 'The document is authentic' is accepted. Conclusions that depend upon the document’s authenticity may then become inferentially available; if that proposition is instead rejected, those same routes can no longer proceed on that basis. Conversely, rejecting the proposition 'The document was altered after its discovery' may remove an obstacle to conclusions that depend upon its evidential reliability. What matters in these cases is not the subject matter of the propositions, but the structural difference produced by their respective orientations. Acceptance and rejection thus do more than identify positive and negative positions for the reason that they determine different possibilities and constraints within the inferential field. Their polarity is therefore operational, since the orientation assigned to a proposition affects what may subsequently be sustained, excluded, or inferentially pursued within logical analysis.

Furthermore, the consequences associated with acceptance and rejection give rise to corresponding inferential commitments. To accept a proposition is not simply to assign it a positive position, but to undertake the inferential commitments associated with that position; similarly, rejecting a proposition carries consequences for what may continue to be maintained within the same analysis. These commitments reveal that inferential positions cannot be understood in isolation, since the position assigned to one proposition may affect the positions available to others. Relations of compatibility, incompatibility, and exclusion are therefore central to the operation of acceptance and rejection. For example, if an investigation accepts the proposition 'The artefact was produced exclusively in the first century' while also accepting reliable evidence that a particular specimen was manufactured in the third century, the resulting positions cannot simply be maintained independently; their incompatibility requires the inferential configuration to be reconsidered. The significance of acceptance and rejection thus extends beyond the positioning of individual propositions to the organisation of relations among them. In this sense, they provide warrant theory with a bilateral means of representing not only inferential orientation, but also the commitments and constraints that arise from the interaction of positive and negative positions within logical analysis.

Taken together, these considerations establish the distinctive role of acceptance and rejection within warrant theory. Their function is to provide a common bilateral vocabulary through which the inferential orientation of propositions, the consequences associated with those orientations, and the commitments and constraints arising from their interaction can be systematically represented. Acceptance and rejection thus connect the inferential entitlement identified by warrant theory with the positive and negative positions examined through warrant-theoretic analysis, without introducing an independent source of warrant. Their importance lies precisely in this organisational role since they enable warrant theory to represent inferential standing not as an undifferentiated form of admissibility, but as a structured field in which propositions occupy opposed orientations whose consequences are systematically related. With this bilateral structure now established, the remaining question concerns the broader conception of logic that follows when warrant, warrant-theoretic analysis, and the positive and negative inferential positions expressed through acceptance and rejection are considered together. It is to this question that the final section now turns.

## 4. A Warrant-Theoretic Definition of Logic

The preceding sections provide the conceptual basis for a warrant-theoretic definition of logic. Warrant has been identified as inferential entitlement, warrant-theoretic analysis has supplied the method for examining the acquisition and development of inferential standing, and acceptance and rejection have distinguished the positive and negative orientations that such standing may assume. When these elements are considered together, logic may be defined as the formal and normative study of the conditions under which propositions may be legitimately accepted or rejected, and of the inferential transitions that such legitimacy authorises within a structured system of analysis. The central feature of this definition is inferential authorisation. Logic, on this account, does not investigate simply which transitions can be formulated or which conclusions happen to be reached; it determines which inferential movements are warranted by the positions already established within an analysis. Its formal character consists in the systematic representation of these relations independently of the particular subject matter under consideration, while its normative character consists in distinguishing warranted transitions from movements for which the required entitlement is absent. Suppose, for example, that a reasoner moves from the proposition 'Every employee attended the meeting' to 'Maria attended the meeting'. That transition becomes warranted only if Maria is established as an employee within the relevant analysis. Without that additional position, the conclusion may happen to be true, but its truth alone does not authorise the inference. The distinction illustrates the specific contribution of the warrant-theoretic definition because logical analysis concerns not simply the occurrence of inferential movement, but the authority under which such movement proceeds. Logic is thus concerned with the systematic conditions governing when inferential positions authorise, require, restrict, or exclude further positions.

This conception permits logical consequence to be understood as a relation of inferential entitlement between positions. When a conclusion follows logically from certain premises, the relevant claim is that the inferential standing of those premises, together with the governing logical principles, authorises the position assigned to the conclusion. Logical consequence is therefore not simply a succession in which one proposition appears after another, but a warranted transition in which the established positions supply sufficient inferential authority for what follows. Consider the familiar →E rule:

$$\frac{\mathbf{A \to B} \qquad \mathbf{[A]}}{\mathbf{B}}\ \mathbf{\to E}$$

Once the inferred conditional A → B and the assumption [A] possess the appropriate positive inferential standing, the conclusion B occupies a position authorised by their conjunction within the relevant inferential structure. By contrast, from A → B and B one cannot infer A simply on that basis. A may nevertheless be true, but its truth would not supply the missing entitlement required for that transition. The contrast demonstrates why a warrant-theoretic account distinguishes a conclusion that happens to be correct from a conclusion that is inferentially warranted. This distinction also reveals the significance of the bilateral framework developed in the preceding section. Inferential authority need not operate solely by permitting further positive positions. An established configuration may instead warrant the rejection of a proposition, prevent a position from being sustained, or render particular combinations inferentially incompatible. Logical consequence can accordingly involve both permission and constraint since it determines not only what may follow from established positions, but also what cannot retain inferential standing in conjunction with them. The bilateral vocabulary of acceptance and rejection thus enables logical consequence to be represented as regulated movement among positive and negative inferential positions, while warrant supplies the entitlement governing that movement.

The same account provides a corresponding explanation of logical failure. If a logical transition requires inferential entitlement, then an inference fails when the position assigned to its conclusion is not authorised by the positions from which it proceeds. Logical error, in this sense, need not consist in the falsity of the resulting proposition; it may instead consist in the absence of a warranted route to that proposition. Suppose that a scientific investigation establishes that a particular substance causes either reaction A or reaction B under specified conditions, and a researcher immediately concludes that reaction A must occur. Even if subsequent observation confirms reaction A, the original transition remains unwarranted because the established disjunction did not authorise the selection of that alternative. The example illustrates an important consequence of the proposed definition namely, the status of an inference depends upon the authority connecting its positions, rather than solely upon the eventual truth of its conclusion. This conception applies equally to extended reasoning. A derivation may contain individually admissible propositions and nevertheless fail if one stage assigns a position that the preceding inferential structure does not warrant. Logical analysis must therefore govern the continuity of inferential entitlement throughout a sequence, identifying both where movement is authorised and where its authority terminates. Within a bilateral framework, this regulatory function also encompasses incompatibility and exclusion. If established positions cannot be jointly sustained, logic must determine what inferential adjustment is required, including whether a proposition must be rejected or whether a proposed transition must be blocked. Logical constraint is consequently as significant as logical permission. The warrant-theoretic conception captures both dimensions within a single framework by treating successful and unsuccessful inference as questions concerning the presence, preservation, or absence of inferential entitlement.

The proposed definition should not be understood as replacing truth, validity, derivability, semantics, or proof theory with a new vocabulary. These notions retain their established roles; warrant theory instead supplies a philosophical standpoint from which their inferential authority may be examined. A semantic framework may determine truth under an interpretation, a proof system may specify rules of derivation, and a formal calculus may establish whether a sequence satisfies the requirements of a particular system.[16] The warrant-theoretic question concerns what these structures permit within logical analysis that is, what permits a proposition to occupy a given inferential position, what further positions that entitlement supports, and where its authority is restricted or exhausted. This also allows the account to remain applicable across different logical systems without assuming that they must employ identical rules or consequence relations. What matters warrant-theoretically is that the relevant system specifies conditions under which inferential positions are warranted and relations through which their standing is governed. The elements developed throughout this manuscript can therefore be brought together without collapsing their distinct functions. Warrant identifies inferential entitlement; warrant-theoretic analysis investigates its operation; and acceptance and rejection articulate the bilateral orientation of the positions in which that entitlement is expressed. Logic concerns the systematic governance of these positions and of the approved transitions and constraints connecting them. Accordingly, the warrant-theoretic definition advanced in this document may be stated in its final form i.e. logic is the formal and normative study of the conditions under which propositions may be legitimately accepted or rejected, and of the inferential transitions that such legitimacy warrants within a structured system of analysis. On this conception, the distinctive philosophical subject matter of logic lies in the systematic governance of inferential legitimacy.

## 5. Concluding Remarks

The aims established at the outset of this paper were to develop warrant theory as a philosophical discipline concerned with inferential legitimacy, to explain warrant as inferential entitlement, to formulate warrant-theoretic analysis, to clarify the bilateral roles of acceptance and rejection, and ultimately to offer a warrant-theoretic definition of logic. The four sections have pursued these aims in a systematic sequence. Section 1 introduced warrant as the entitlement that licenses a proposition to function within logical analysis, distinguishing warrant from truth, belief, assertion, and psychological conviction, while connecting inferential use with meaning. It also showed, through the introduction and elimination rules of natural deduction, how warrant may be generated, transmitted, and preserved. Section 2 transformed this conceptual account into a method namely, warrant-theoretic analysis investigates how propositions acquire inferential standing, how that standing is maintained, transferred, restricted, or lost, and how inferential positions interact through dependence, compatibility, incompatibility, and exclusion. Section 3 supplied the bilateral vocabulary required to articulate the direction of such positions. Acceptance was characterised as positive inferential positioning and rejection as negative inferential positioning, while both were distinguished from speech acts and psychological attitudes. Section 4 then brought these components together by defining logic as the formal and normative study of the conditions governing legitimate acceptance and rejection and the inferential transitions warranted by established positions. Logical consequence and logical failure were consequently interpreted through the presence, preservation, or absence of inferential entitlement.

[16] For a concise but technical understanding of a proof system, see Irani (2025) Page: 789.

The account developed here is intended as a foundation rather than an exhaustive theory, and warrant theory consequently opens several directions for future research. One important task is to formulate a warrant-theoretic theory of logical consequence. Although the present discussion has characterised logical consequence in terms of inferential entitlement between positions, further work can determine systematically the conditions under which warrant is preserved across consequence relations, how positive and negative positions contribute to consequence, and how different logical systems regulate such preservation. A second central project is to provide a warrant-theoretic definition of the inferential meaning of logical constants. The discussion of Gentzen and the harmony between introduction and elimination rules suggests that the meaning of a logical constant may be investigated through the conditions under which propositions containing it acquire warrant and through the consequences their warranted use authorises. Developing this proposal could provide the basis for the warrant-theoretic semantics identified in this paper. Further research may also investigate the relations between warrant and proof-theoretic validity, the treatment of incompatibility and exclusion in bilateral systems, and the conditions governing the acquisition, defeat, restoration, or transfer of warrant across extended derivations. Beyond formal logic, warrant theory may be applied to questions in the philosophy of language, epistemology, ethics, and other normative domains in which inferential entitlement plays an organising role. These projects would test the scope and explanatory power of the framework while preserving its central commitment since logic concerns the systematic governance of inferential legitimacy and the authority under which propositions occupy and move between inferential positions.